\documentclass[12pt]{article}

\usepackage{epsfig,amsmath,amssymb} 
\usepackage{graphics}

\newcommand{\be}{\begin{equation}}
\newcommand{\ee}{\end{equation}}
\newcommand{\bi}[1]{\vspace{-3mm} \bibitem{#1}}

\begin{document}

\begin{center}
Communications in Nonlinear Science and Numerical Simulation 11 (2006) 885-898
\end{center}

\begin{center}
{\Large \bf 
Fractional dynamics of systems with long-range interaction }
\vskip 5 mm

{\large \bf Vasily E. Tarasov$^{1,2}$ and George M. Zaslavsky$^{2,3}$ } \\

\vskip 3mm

{\it $1)$ Skobeltsyn Institute of Nuclear Physics, \\
Moscow State University, Moscow 119992, Russia } \\
{\it $2)$ Courant Institute of Mathematical Sciences, New York University \\
251 Mercer Street, New York, NY 10012, USA, }\\ 
{\it $3)$ Department of Physics, New York University, \\
2-4 Washington Place, New York, NY 10003, USA } \\
\end{center}

\vskip 11 mm

\begin{abstract}
We consider one-dimensional chain of coupled linear and nonlinear
oscillators with long-range power wise interaction defined by a term
proportional to $1/|n-m|^{\alpha+1}$. 
Continuous medium equation for this system can be obtained in
the so-called infrared limit when the wave number tends to zero. 
We construct a  transform operator
that maps the system of large number of ordinary differential equations 
of motion of the particles into a partial differential equation 
with the Riesz fractional derivative of order $\alpha$, when 
$0<\alpha<2$. Few models of coupled oscillators are considered and  
their synchronized states and localized structures are discussed in 
details. Particularly, we discuss some solutions of time-dependent 
fractional Ginzburg-Landau (or nonlinear Schrodinger) equation.
\end{abstract}

\vskip 3 mm
{\small 



\vskip 3 mm


\vskip 11 mm

\section{Introduction}

Although the fractional calculus is known for more than two
hundred years and its developing is an active area of mathematics,
appearance and use of it in physical literature
is fairly recent and sometimes is considered as exotic.
In fact, there are many different areas where fractional
equations, i.e., equations with fractional integro-differentiation,
describe real processes.
Between the most related areas are chaotic dynamics
\cite{Zaslavsky1}, random walk in fractal space-time \cite{Montr}
and random processes of Levy type \cite{SZ,Uch,MS1,MS2}.
The physical reasons for the appearance of fractional equations
are intermittancy, dissipation, wave propagation in complex
media, long memory, and others. An important area of the application of
long-range interaction includes collective oscillation and 
synchronization in physics, chemistry, biology, and neuroscience
actively studied recently \cite{Afr,Pik1,BKOVZ}. 
Beginning from the pioneering contributions
by Winfree \cite{Win} and Kuramoto \cite{Kur1},
studies of synchronization in populations of coupled oscillators
becomes an active field of research in biology and chemistry.
The synchronization can also be considered in complex
oscillatory medium, where each site (element)
performs self-sustained oscillations.
A good physical and chemical example is the
oscillatory Belousov-Zhabotinsky reaction
\cite{Belousov,Zhab,Kur1} in a medium where different
sites can oscillate with different periods and phases.

Complex Ginzburg-Landau equation is canonical model for
oscillatory systems with local coupling near Hopf bifurcation.
Recently, Tanaka and Kuramoto \cite{TK} have shown how, in the
vicinity of the bifurcation, the description of an array of nonlocally
coupled oscillators can be reduced to the complex Ginzburg-Landau
equation.
In Ref. \cite{Mikh}, a model of population of diffusively
coupled oscillators with limit-cycles is described by the complex
Ginzburg-Landau equation with nonlocal interaction.

Nonlocal coupling means interaction with oscillators or other
objects distanced arbitrary far from each others. The
corresponding interaction potential could be with a
characteristic length  Refs. \cite{Kur3,Kur4,Mikh} or without it.
The long-range interaction that decreases as $1/|x|^{\alpha+1}$
with $0< \alpha <2$ was considered in Refs.
\cite{Dyson,J,CMP,NakTak,S} with respect to the system's
thermodynamics and phase transition.
It is also shown in \cite{Lask} that using the Fourier transform
and limit for the wave number $k \rightarrow 0$, 
the long-range term interaction
leads under special conditions to the fractional dynamics.

In the last decade it is found that many physical processes can
be adequately described by equations that consist of derivatives 
of fractional order.
In a fairly short period of time the list of such
applications becomes long and the area of applications is broad.
Even in a concise form, the applications include material
sciences \cite{Hilfer,C2,Nig1,Nig3}, chaotic dynamics \cite{Zaslavsky1},
quantum theory \cite{Laskin,Naber,Krisch},
physical kinetics \cite{Zaslavsky1,SZ,ZE,Zaslavsky7},
fluids and plasma physics \cite{CLZ,Plasma2005},
and many others physical topics related 
to wave propagation \cite{ZL}, long-range dissipation \cite{GM},
anomalous diffusion and transport theory
(see reviews \cite{Zaslavsky1,Montr,Hilfer,Uch,MK}).
Particularly, fractional Ginzburg-Landau equation was suggested
in \cite{Zaslavsky6,Physica2005,Mil}.

In this paper we would like to strength the state that the 
one-dimentional chain of particles with long-range power type
interaction  can be asymptotically described by the continuous
medium equation with fractional space derivative. 
We introduce, under some conditions,  a corresponding transform
operator that map the system of large number ordinary
differential equations of particles motion into the equation
with partial and fractional derivatives. It is worthwhile to
mention that possibility of such an operation has an immediate physical
consequence: fractional  dynamics equation has solutions 
that describe coherent structures with power-like tails.

In  Section 2,  we introduce the transform operator. 
In Section 3, some particular solutions are derived with
a constant wave number for the fractional Ginzburg-Landau (FGL)
equation that can be interpreted as
synchronization in the oscillatory medium.
In Section 4, we consider forced FGL equation for
isochronous case.
In Section 5, we derive solutions of the fractional
Ginzburg-Landau equation  near a limit cycle.
These solutions are interpreted as coherent structures in
the oscillatory medium with long-range interaction.

\section{Long-range interaction of oscillators}

Consider a system with Hamiltonian
\be \label{Z1}
H =  \sum_{n=-\infty}^{+\infty} \left[ 
\frac{1}{2} \; \dot{u}_n^2 + 
\frac{1}{2}g_0 \sum_{\substack{m=-\infty \\ m \ne n}}^{+\infty} \; 
\frac{1}{|n-m|^{1+\alpha}} \; u_n u_m + \; V (u_n)  \right],
\ee
where $g_0$ and $\alpha$ are some constants, and $u_n=u_n(t)$. 
The corresponding equations of motion are
\be \label{Z2a}
\frac{\partial^2 u_n }{\partial t^2}+
g_0 \sum_{\substack{m=-\infty \\ m \ne n}}^{+\infty} \; 
\frac{1}{|n-m|^{1+\alpha}} \; u_m + \; V^{\prime} (u_n)=0 ,
\ee
with $V^{\prime} (u)=\partial V (u)/ \partial u$.
Let us introduce notations
\be \label{Z2b}
H_{int} \equiv \frac{1}{2} g_0 
\sum_{\substack{m=-\infty \\ m \ne n}}^{+\infty} \; 
|n-m|^{-(1+\alpha)} \; u_n u_m ,
\ee
and
\be \label{Z3}
\hat T_{\alpha} u_n \equiv 
\sum_{\substack{m=-\infty \\ m \ne n}}^{+\infty} \; 
|n-m|^{-(1+\alpha)} \; u_m .
\ee
The standard transformation from the set of equations (\ref{Z2a}) 
to the continuous medium equations is in replacements
\be \label{Z4}
u_n(t) \equiv u(x_n,t) \rightarrow u(x,t), \quad x_n=n \Delta x,
\ee
where $n$ is a positive integer number, $\Delta x$ is a distance
between particles, and the limit $\Delta x \rightarrow 0$ 
is applied.

A specific form of $H_{int} $ and the 
corresponding term $\hat T_{\alpha}$ in the equation of motion 
(\ref{Z2a}) create a possibility to present the continuous medium equations
in a form that consists of fractional derivatives.
For this goal, we use the notation
\[
u(x,t) = \frac{1}{2\pi} \int_{-\infty}^{+\infty} dk \; \tilde u(k,t) \; 
e^{ikx} \equiv \mathcal{F}^{-1}\{\tilde u(k,t)\} ,
\]
\be \label{Z5}
\tilde u(k,t)= \int_{-\infty}^{+\infty} dx \; Z(x,t) \; 
e^{-ikx} \equiv \mathcal{F}\{u(x,t)\} ,
\ee
where $\mathcal{F}\{u(x,t)\}$ is Fourier transform of $u(x,t)$ 
with respect to $x$.
Then, replacing (\ref{Z4}) in (\ref{Z2a}), we obtain
$\hat T_{\alpha}u_n(t) \rightarrow \hat T_{\alpha} u(x,t)$,
and after applying Fourier transform to (\ref{Z2a}),
\be \label{Z6}
\frac{\partial^2 u(k,t) }{\partial t^2}+
g_0 [T_{\alpha}(k)-T_{\alpha}(0)] \; u(k,t) + 
\; \mathcal{F} \{V^{\prime} (u(x,t))\}=0 ,
\ee 
where
\be \label{Z7}
T_{\alpha}(k)=2\Gamma(-\alpha)\cos(\pi \alpha/2) |k|^{\alpha}+
2\sum^{\infty}_{n=0} \frac{\zeta(\alpha+1-2n)}{(2n)!} (-k^2)^n , 
\quad |k|< 1, \quad \alpha \not=0,1,2,3...,
\ee
and  $T_{\alpha}(0)=2 \zeta(1+\alpha)$ and $\zeta$ 
is the Riemann zeta-function.
Function $T_{\alpha}(k)$ can also be presented in the form
\be \label{Z8}
T_{\alpha}(k)=
\sum_{\substack{n=-\infty \\ n \ne 0}}^{+\infty} 
e^{-ikn} \frac{1}{|n|^{\alpha+1}}
=2\sum^{\infty}_{n=1} \frac{\cos(kn)}{n^{1+\alpha}} 
=Li_{\alpha+1}( e^{ik} ) +Li_{\alpha+1}( e^{-ik} ) ,
\ee
where $Li_{\alpha}(z)$ is a polylogarithm function \cite{Erd}
\be \label{Z9}
Li_{\beta}(e^z)=\Gamma(1-\beta) (-z)^{\beta-1}+\sum^{\infty}_{n=0}
\frac{\zeta(\beta-n)}{n!} z^n, \quad |z|< 2\pi , \quad
\beta \not= 1,2,3....
\ee

After substitution (\ref{Z7}) into (\ref{Z6}) obtain
\be \label{D4}
\frac{\partial^2 \tilde u(k,t)}{\partial t^2}=
\mathcal{F}\{V^{\prime}(u(x,t))\}-
g_0 a_{\alpha} |k|^{\alpha} \tilde u(k,t)-
2g_0 \sum^{\infty}_{n=1} \frac{\zeta(\alpha+1-2n)}{(2n)!} (-k^2)^n \tilde u(k,t),
\ee
where 
\be \label{D5}
a_{\alpha} =2\Gamma(-\alpha)\cos(\pi \alpha/2) , 
\quad \alpha \not= 0,1,2,... .
\ee
From Eq. (\ref{D4}) we obtain the equation for the field $u(x,t)$ 
using the definition (\ref{Z5})
\be \label{D7}
\frac{\partial^2 u(x,t)}{\partial t^2}= V^{\prime}(u(x,t))+
g_0 a_{\alpha} \frac{\partial^{\alpha}}{\partial |x|^{\alpha}} u(x,t) -
2g_0 \sum^{\infty}_{n=1} \frac{\zeta(\alpha+1-2n)}{(2n)!} 
\frac{\partial^{2n}}{\partial x^{2n}} u(x,t) ,  
\quad (\alpha \not=0,1,2...) .
\ee
Here, we use the connection between Riesz fractional derivative 
and its Fourier transform \cite{SKM}: 
\be
|k|^{\alpha} \longleftrightarrow
- \frac{\partial^{\alpha}}{\partial |x|^{\alpha}} ,
\quad
k^2 \longleftrightarrow
- \frac{\partial^2}{\partial |x|^2} .
\ee
Such construction that leads to the equation with fractional derivative
appears also in \cite{Lask}.

The properties of the Riesz derivative can be found in 
\cite{SKM,OS,Podlubny}. Its another expression is
\be \label{R1}
\frac{\partial^{\alpha}}{\partial |x|^{\alpha}} u(x,t)=
-\frac{1}{2 \cos(\pi \alpha /2)} 
\left({\cal D}^{\alpha}_{+}u(x,t) +{\cal D}^{\alpha}_{-} u(x,t)\right) ,
\ee
where $\alpha\not=1,3,5...$, and
${\cal D}^{\alpha}_{\pm}$ are Riemann-Liouville 
left and right fractional derivatives
\[
{\cal D}^{\alpha}_{+}u(x,t)=
\frac{1}{\Gamma(n-\alpha)} \frac{\partial^n}{\partial x^n}
\int^{x}_{-\infty} \frac{u(\xi,t) d\xi}{(x-\xi)^{\alpha-n+1}},
\]
\be \label{R3}
{\cal D}^{\alpha}_{-}u(x,t)=
\frac{(-1)^n}{\Gamma(n-\alpha)} \frac{\partial^n}{\partial x^n}
\int^{\infty}_x \frac{u(\xi,t) d\xi}{(\xi-x)^{\alpha-n+1}} ,
\ee
where $n-1< \alpha <n$.

The obtained expressions (\ref{D4}) and (\ref{D7}) can be simplified in
the so-called infrared approximation $k \rightarrow 0$. 
Then Eq.\ (\ref{D4}) can be rewritten as
\be \label{D7b}
\frac{\partial}{\partial t}\tilde u(k,t)=
\mathcal{F}\{V^{\prime}(u(x,t))\}+
g_0 \mathcal{T}_{\alpha}(k)  \tilde u(k,t), 
\quad k \rightarrow 0, \quad (1<\alpha<3) ,
\ee
where we use $\mathcal{T}_{\alpha}(k)$ as an approximation 
expression for $T_{\alpha}(k)$:
\be \label{D8}
\mathcal{T}_{\alpha}(k) 
= 2 \Gamma (-\alpha) \cos (\pi \alpha/2) |k|^{\alpha}- 
\zeta (\alpha -1) k^2, \; \; \; (1<\alpha<2), 
\ee
\[
\mathcal{T}_{2}(k) = const \; \ln k^{2}, \; \; \; (\alpha=2), 
\]
\[
\mathcal{T}_{\alpha}(k) = J_0 k^{2}, \; \; \; (\alpha>2).
\]
The expression for $\mathcal{T}_{\alpha}(k)$
can be considered as a Fourier transform 
of an operator $\hat{\mathcal{T}_{\alpha}}(x)$:
\be \label{hatTx}
\hat{ \mathcal{T}_{\alpha}}(x) =
\mathcal{F}^{-1} \{ \mathcal{T}_{\alpha} (k) \} = 
a_{\alpha} \frac{\partial^{\alpha}}{\partial |x|^{\alpha}} +
 \zeta (\alpha -1) \frac{\partial^2}{\partial x^2}.
\ee
Applying inverse Fourier transform to (\ref{D7b}), we also obtain
\be \label{D10}
\frac{\partial^2}{\partial t^2}u(x,t)=
V^{\prime}(u(x,t))+g_0 \hat{ \mathcal{T}_{\alpha}}(x) u(x,t) ,
\quad (1<\alpha<2) . 
\ee
This equation can be considered as an equation for continuous 
oscillatory medium with $1<\alpha<2$ in the infrared ($k \rightarrow 0$) 
approximation.  

Let us note that (\ref{D8}) has a scale $l_0$: 
\be 
l^{-1}_0 \equiv |2 \Gamma (-\alpha) 
\cos (\pi \alpha/2)/\zeta (\alpha-1)|^{1/(2-\alpha)}
\ee
such that nontrivial expression 
$\mathcal{T}_{\alpha} (k) \sim |k|^{\alpha}$ appears for $k \ll k_0$. 
In the following, we consider the case $k \ll k_0$, i.e.
\be \label{Z25}
\hat{ \mathcal{T}_{\alpha}}(x) \sim 
a_{\alpha} \frac{\partial^2}{\partial |x|^{\alpha}}.
\ee
The expression (\ref{Z7}) for $T_{\alpha}(k)$ maps the system (\ref{Z2a})
for large number of oscillators into the continuous medium
equation (\ref{Z6}) in the limit (\ref{Z4}).
The approximate expression $\mathcal{T}_{\alpha} (k)$  (\ref{D8})
for $T_{\alpha}(k)$ is used to construct an operator 
$\hat{ \mathcal{T}_{\alpha}} (x)$ that stands instead of 
$\hat T_{\alpha}$ in (\ref{Z3}) and that appears in 
the final approximate equation (\ref{D10}).
We will call $\hat{ \mathcal{T}_{\alpha}} (x)$ the transform operator.
The reason for that is that it can be used for in more broad
sense than just for the Hamiltonian (\ref{Z1}). 
Here are few examples.

The dynamical equations can be gain the term of the type (\ref{Z2b})
in complex dispersive media \cite{Nig1,ZL,Zaslavsky7}.
The corresponding approximate equations with fractional derivatives
appear for tracer dynamics in the presence of convective rolls \cite{YPP}, 
and for the equation for surface wave interaction in ocean \cite{MMT}.
A specific type of equations is widely used for the oscillatory media:
\be \label{C1}
\frac{d}{dt} z_n(t)= F(z_n)+g_{0}             
\sum_{\substack{m=-\infty \\ m \ne n}}^{+\infty} 
J_{\alpha} (n-m) (z_n-z_m) ,
\ee
where $z_n$ is the position of the $n$th oscillator in the complex plane,
and $F$ is a force. As an example, for the oscillators 
with a limit cycle, $F$ can be taken as
\be \label{E2}
F(z)=(1+ia)z-(1+ib) |z|^2 z .
\ee
The nonlocal interaction is given by the power function
\be \label{C2}
J_{\alpha}(n)= |n|^{-\alpha-1} .
\ee
Similarly to (\ref{Z3}). Using the notations (\ref{D8}) 
and (\ref{hatTx}), we can rewrite the approximate equation
with fractional derivative in the form
\be
\frac{\partial Z(x,t)}{\partial t}=F(Z(x,t))+
g_0 \hat{ \mathcal{T}_{\alpha}}(x) Z(x,t), \quad (1<\alpha<2) ,
\ee
where we replace $z_n(t) \rightarrow Z(x,t)$ and 
consider $k \rightarrow 0$.

Using, for example $F(z)=0$ and expression (\ref{Z25}) for 
$\hat{ \mathcal{T}_{\alpha}} (x)$, we arrive to 
the fractional kinetic equation: 
\be \label{D10b}
\frac{\partial}{\partial t}Z(x,t)=g_0 a_{\alpha} 
\frac{\partial^{\alpha}}{\partial |x|^{\alpha}} Z(x,t) ,
\quad (1<\alpha<2) 
\ee
that describes the fractional superdiffusion \cite{SZ,Zaslavsky7,Uch}.
For $F(z)$ in the form (\ref{E2}), we obtain
fractional Ginzburg-Landau equation suggested in 
\cite{Zaslavsky6,Physica2005,Mil} that will be 
considered in the next section.

Finally, consider
\be
H_{int}= \frac{1}{2} g_0 
\sum_{\substack{n,m=-\infty \\ n \ne m}}^{+\infty} 
\frac{1}{|n-m|^{1+\alpha}} g(u_n) \; g(u_m) ,
\ee
where $g(u)$ is some function of $u$, instead of (\ref{Z2b}). 
The corresponding generalization of (\ref{D10})
is easily obtained as 
\be
\frac{\partial^2 u(x,t)}{\partial t^2} = V^{\prime}(u(x,t))+
g_0 \; g^{\prime}(u(x,t)) \hat{ \mathcal{T}_{\alpha}}(x) \; g(u(x,t)), 
\; \; k \rightarrow 0, \; \; (1< \alpha <2),
\ee
where $g^{\prime}(u)=\partial g(u)/\partial u$ and 
$\hat{ \mathcal{T}_{\alpha}}(x)$ is the 
same as in (\ref{hatTx}) or (\ref{Z25}).

\section{Fractional Ginzburg-Landau (FGL) equation}

The one-dimensional lattice of weakly coupled nonlinear 
oscillators is described by 
\be
\frac{d}{dt} z_n(t)=(1+ia)z_n -(1+ib)|z_n|^2z_n+
(c_1+ic_2)(z_{n+1}-2z_{n}+z_{n-1}),
\ee
where we assume that all oscillators have the same parameters.
A transition to the continuous medium assumes \cite{Pik1}
that the difference $z_{n+1}-z_n$ is of order $\Delta x$, and
the interaction constants $c_1$ and $c_2$ are large.
Setting $c_1=g (\Delta x)^{-2}$, and $c_2=gc(\Delta x)^{-2}$,
we get
\be \label{A1}
\frac{\partial}{\partial t}Z=(1+ia)Z-(1+ib)|Z|^2Z+
g(1+ic)\frac{\partial^2}{\partial x^2} Z ,
\ee
which is a complex time-dependent Ginzburg-Landau equation.
The simplest coherent structures for this equation are 
plane-wave solutions \cite{Pik1}:
\be \label{A3}
Z(x,t)=R(K)exp[ iKx-i\omega(K)t+\theta_0],
\ee
where
\be \label{A4}
R(K)=(1-gK^2)^{1/2} , \quad 
\omega(K)=(b-a)+(c-b)gK^2 ,
\ee 
and $\theta_0$ is an arbitrary constant phase.
These solutions exist for $gK^2<1$. 
Solution (\ref{A3}) can be interpreted as a synchronized state \cite{Pik1}.

Let us come back to the equation for
nonlinear oscillators (\ref{C1}) with $F(z)$ in Eq. (\ref{E2}) 
and long-range coupling (\ref{C2}):
\be \label{D11}
\frac{d}{dt}z_n=(1+ia)z_n-(1+ib)|z_n|^2 z_n +
g_0 \sum_{m \not=n} \frac{1}{|n-m|^{\alpha+1}} (z_n-z_m) ,
\ee
where $z_n=z_n(t)$ is the position of the $n$th oscillator 
in the complex plane, $1<\alpha<2$.
The corresponding equation in the continuous limit and 
infrared approximation can be obtained in the 
same way as (\ref{D10}):
\be \label{A2}
\frac{\partial}{\partial t}Z=(1+ia)Z-(1+ib)|Z|^2Z+
g(1+ic)\frac{\partial^{\alpha}}{\partial |x|^{\alpha}} Z ,
\ee
where $g(1+ic)=g_0 a_{\alpha}$, and $1<\alpha<2$.
Eq. (\ref{A2}) is a fractional generalization of complex 
time-dependent Ginzburg-Landau equation (\ref{A1}) (compare to (\ref{D10})). 
Here, this equation is derived in a specific approximation 
for the oscillatory medium.

We seek a particular solution of (\ref{A2}) in the form 
\be \label{A6}
Z(x,t)=A(K,t)e^{iKx} .
\ee
Eq. (\ref{A6}) represents a particular solution of (\ref{A2}) 
with a fixed wave number $K$.

Substitution of (\ref{A6}) into (\ref{A2}) gives
\be \label{A8}
\frac{\partial}{\partial t}A(K,t)=(1+ia)A-(1+ib)|A|^2A-
g(1+ic)|K|^{\alpha}A .
\ee
Rewriting this equation in polar coordinates
\be \label{A9}
A(K,t)=R(K,t)e^{i\theta(K,t)},
\ee
we obtain 
\be \label{A10}
\frac{dR}{dt}= (1-g|K|^{\alpha})R-R^3 ,
\quad
\frac{d \theta}{dt}= (a-cg|K|^{\alpha})-b R^2.
\ee
The limit cycle here is a circle with the radius
\be \label{A11}
R=(1-g|K|^{\alpha})^{1/2} , \quad g|K|^{\alpha} <1 .
\ee
Solution of (\ref{A10}) with arbitrary initial conditions
$R(K,0)=R_0$, $\theta(K,0)=\theta_0$ is
\be \label{B2}
R(t)=R_0 (1-g|K|^{\alpha})^{1/2} 
\left( R^2_0+(1-g|K|^{\alpha}-R^2_0)e^{-2(1-g|K|^{\alpha})t} \right)^{-1/2},
\ee
\be \label{B3}
\theta(t)=
-\frac{b}{2} \ln \left[ (1-g|K|^{\alpha})^{-1} 
\left( R^2_0 +(1-g|K|^{\alpha}-R^2_0)e^{-2at} \right) \right]
-\omega_{\alpha}(K) t+\theta_0 ,
\ee
where
\be \label{B7}
\omega_{\alpha}(K)=(b-a)+(c-b)g|K|^{\alpha} ,
\quad 1-g|K|^{\alpha}>0 . 
\ee
This solution can be interpreted as a coherent structure in 
nonlinear oscillatory medium with long-range interaction.

If $R^2_0=1-g|K|^{\alpha}$,  $g|K|^{\alpha} <1$,
then Eqs. (\ref{B2}) and (\ref{B3}) give
\be \label{Rtheta}
R(t)=R_0, \quad \theta(t)=-\omega_{\alpha}(K) t +\theta_0 .
\ee
Solution (\ref{Rtheta}) means that
on the limit cycle (\ref{A11}) the angle variable $\theta$
rotates with a constant velocity $\omega_{\alpha}(K)$.
As the result, we have the plane-wave solution
\be \label{B6}
Z(x,t)=(1-g|K|^{\alpha})^{1/2} e^{iKx-i\omega_{\alpha}(K)t+i\theta_0} , 
\quad 1-g|K|^{\alpha}>0 ,
\ee
which can be interpreted as synchronized state
of the oscillatory medium.

The plane-wave solution (\ref{B6})
is stable if parameters $a$, $b$, $c$ and $g$ satisfy 
\be \label{Stab}
0<1-g|K|^{\alpha} < a/b-(c/b)g |K|^{\alpha} < 3(1-g|K|^{\alpha}) .
\ee
Condition (\ref{Stab}) defines the region of parameters 
for plane waves where the synchronization exists.

For initial amplitude that deviates from (\ref{A11}), i.e.,
$R^2_0 \not= 1-g|K|^{\alpha}$, 
an additional phase shift occurs due to the term which is
proportional to $b$ in (\ref{B3}).
The oscillatory medium can be characterized by a single generalized 
phase variable.
To define it, let us rewrite (\ref{A10}) as
\be \label{A10b}
\frac{d}{dt} \ln R= (1-g|K|^{\alpha})-R^2 ,
\quad
\frac{d}{dt} \theta= (a-cg|K|^{\alpha})-b R^2.
\ee
From (\ref{A10b}), we obtain
\be \label{54}
\frac{d}{dt} \phi= - \omega_{\alpha}(K) .
\ee
where
\be \label{A10d}
\phi(R,\theta)=\theta-b \ln R 
\ee
is the generalized phase \cite{Pik1}, 
and $\omega_{\alpha}(K)$ is defined by (\ref{B7}).

Eq. (\ref{54}) means that generalized phase $\phi(R,\theta)$ rotates uniformly 
with constant velocity.
For $g|K|^{\alpha}=(b-a)/(b-c)<1$, we have the lines of 
constant generalized phase.
On $(R,\theta)$ plane these lines are logarithmic 
spirals $\theta-b \ln R =const$.
The decrease of $\alpha$ corresponds to the increase of $K$.
For the case $b=0$ instead of spirals we have straight lines $\phi=\theta$.

Energy propagation can be characterized by
the group velocity 
$v_{\alpha,g}={\partial \omega_{\alpha}(K)}/{\partial K}$.
From Eq. (\ref{B7}), we obtain
\be
v_{\alpha,g}=\alpha (c-b) g |K|^{\alpha-1}. 
\ee
For
$|K|<K_1=\left(\alpha / 2 \right)^{2-\alpha}$,
we get 
$|v_{\alpha,g}| > | v_{2,g}|$.
The phase velocity is 
\be
v_{\alpha,ph}=\omega_{\alpha}(K) / K=(c-b)g |K|^{\alpha-1} .
\ee
For $|K|<K_2=2^{\alpha-2}$,
we have
$|v_{\alpha,ph}| > |v_{2,ph}|$.
Therefore long-range interaction decreasing as $|x|^{-(\alpha+1)}$ with
$1<\alpha<2$ leads to increase the group
and phase velocities for small wave numbers ($K \rightarrow 0$).
Note that the ratio $v_{\alpha,g} /v_{\alpha,ph}$ between
the group and phase velocities of plane waves is equal to $\alpha$.

\section{Forced FGL equation for isochronous case}

In this section, we consider FGL equation (\ref{A8})
forced by a constant $E$ (the so-called
forced isochronous case ($b=0$) \cite{Pik1}):
\be \label{A8e}
\frac{\partial}{\partial t}A=(1+ia)A-|A|^2A-
g(1+ic)|K|^{\alpha}A -iE, \quad (\mathrm{Im} E=0),
\ee
where $A=A(K,t)$, and we put for simplicity $b=0$, and $K$ 
is a fixed wave number. 
Our main goal will be transition to a synchronized states and 
its dependence on the order $\alpha$ of the long-range interaction.

Numerical solution of Eq. (\ref{A8e}) was performed with parameters
$a=1$, $g=1$, $c=70$, $E=0.9$, $K=0.1$, 
for $\alpha$ within interval $\alpha \in(1;2)$.
The results are presented on Fig. 1.
For $\alpha_0< \alpha<2$, where $\alpha_0\approx 1.51...$,  
the only stable solution is a stable fixed point.
This is region of perfect synchronization (phase locking), 
where the synchronous oscillations have a constant amplitude
and a constant phase shift with respect to the external force.
For $\alpha<\alpha_0$ the global attractor for (\ref{A8e})
is a limit cycle.
Here, the motion of the forced system is quasiperiodic.
For $\alpha=2$ there is a stable node.
When $\alpha$ decreases, the stable mode transfers into
a stable focus.
At the transition point it loses stability, and 
a stable limit cycle appears.
As the result, we have that the decrease of 
order $\alpha$ from 2 to 1 
leads to the loss of synchronization (see Figure 1).
For $\alpha=2.00$, and $\alpha=1.60$,
we see that in the synchronization region
all trajectories are attracted to a stable node.

In Fig. 1, for $\alpha=2.00$, and $\alpha=1.60$,
we see that in the synchronization region
all trajectories are attracted to a stable node.
For $\alpha=1.54$, $\alpha=1.52$, $\alpha=1.50$, we see that
a stable limit cycle appears via the Hopf bifurcation.
For $\alpha=1.54$, and $\alpha=1.52$, near the boundary of 
synchronization the fixed point is a focus.
For $\alpha=1.50$, and $\alpha=1.4$, the amplitude of the limit cycle grows, 
and synchronization breaks down.


\begin{figure}
\centering
\rotatebox{270}{\includegraphics[width=6.7 cm,height=6.7 cm]{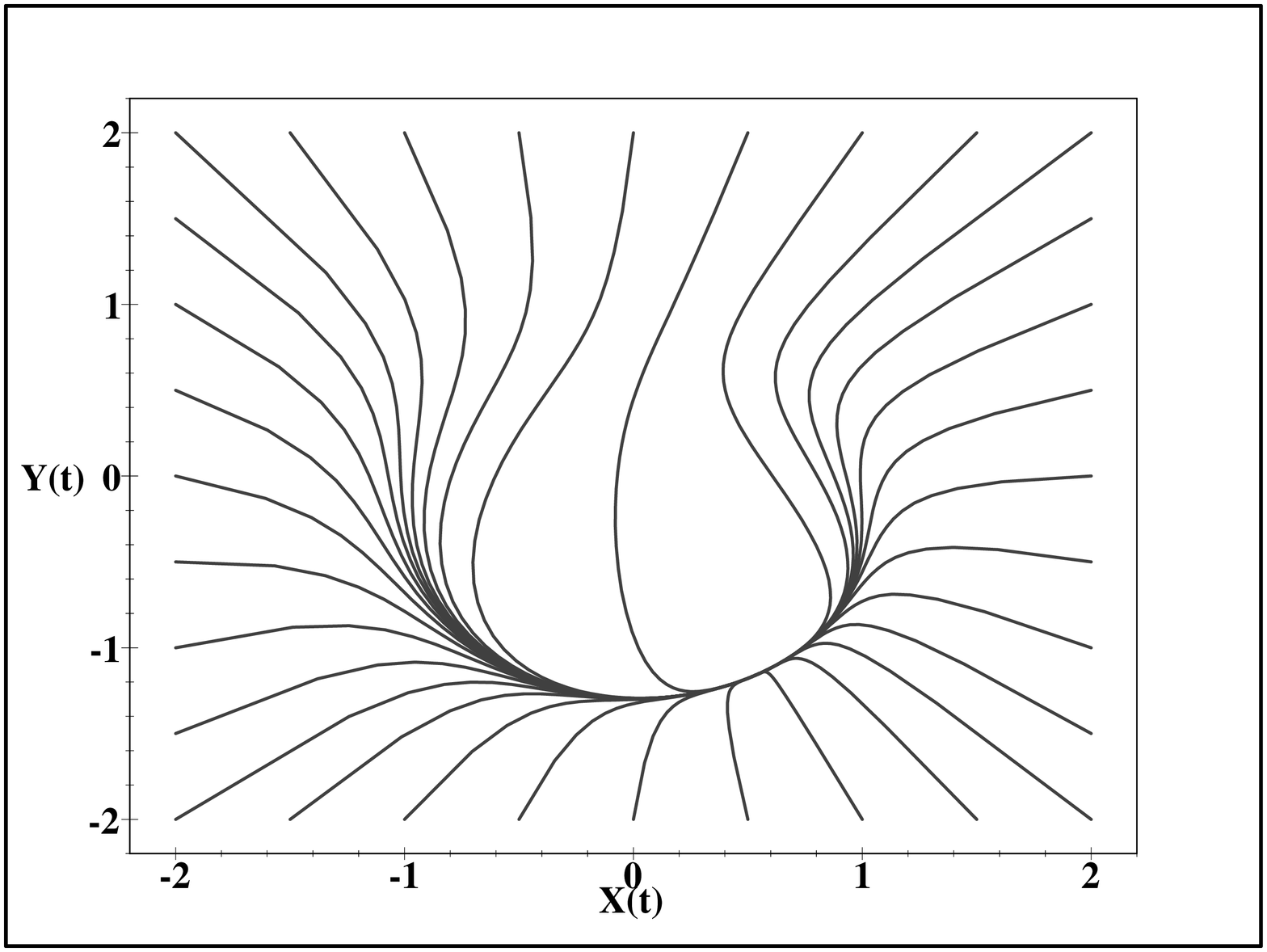}}
\rotatebox{270}{\includegraphics[width=6.7 cm,height=6.7 cm]{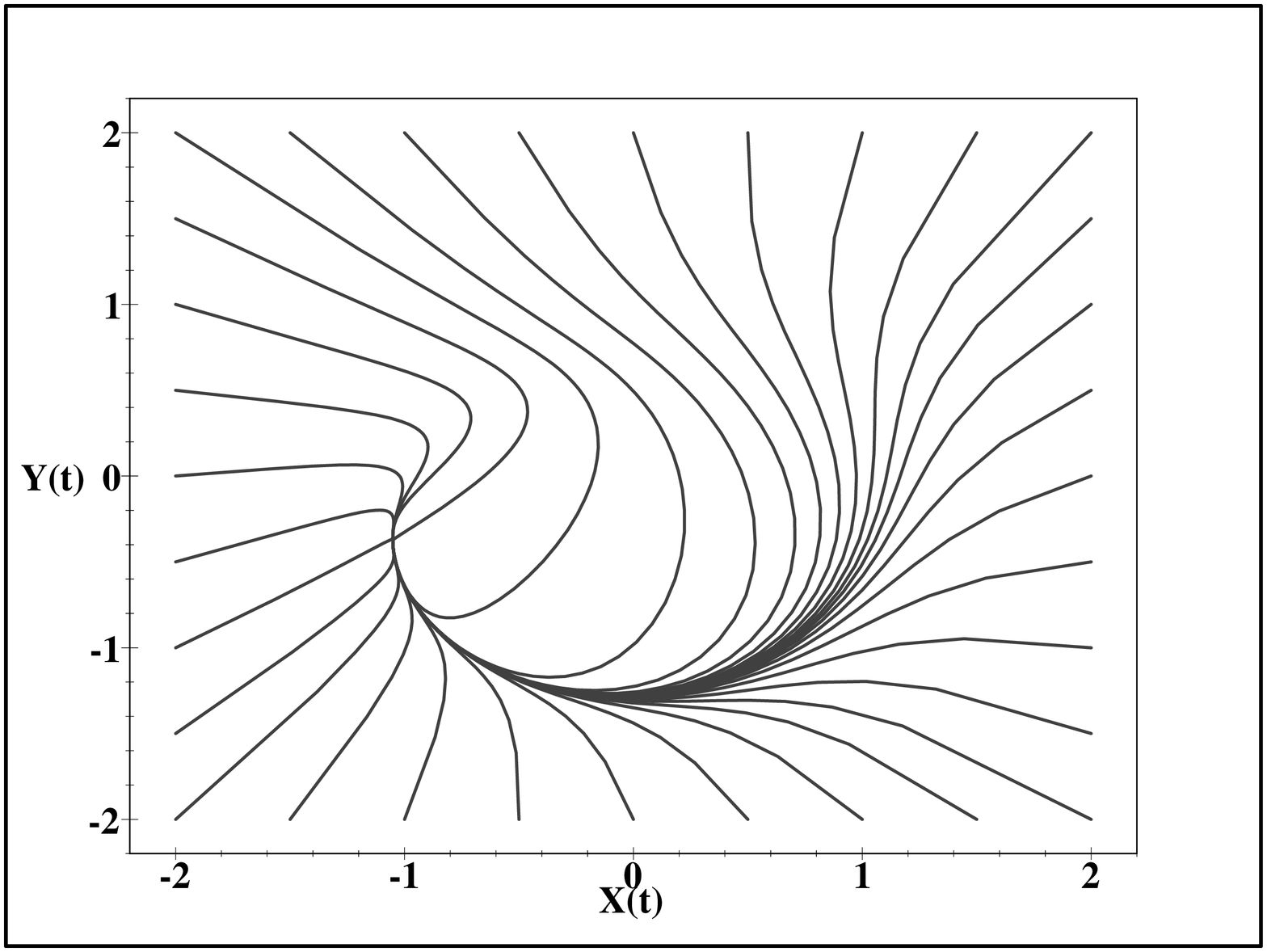}}
\rotatebox{270}{\includegraphics[width=6.7 cm,height=6.7 cm]{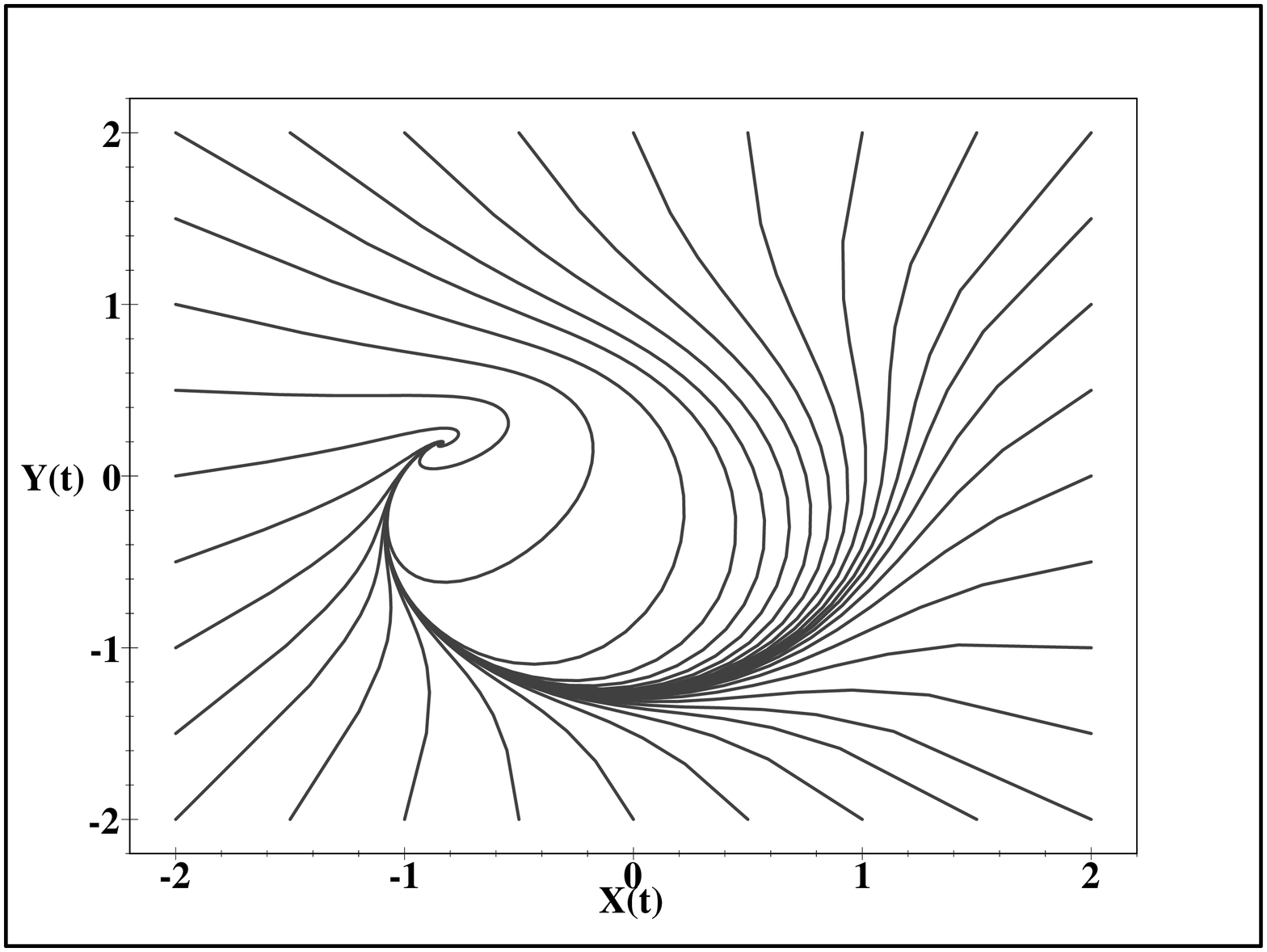}}
\rotatebox{270}{\includegraphics[width=6.7 cm,height=6.7 cm]{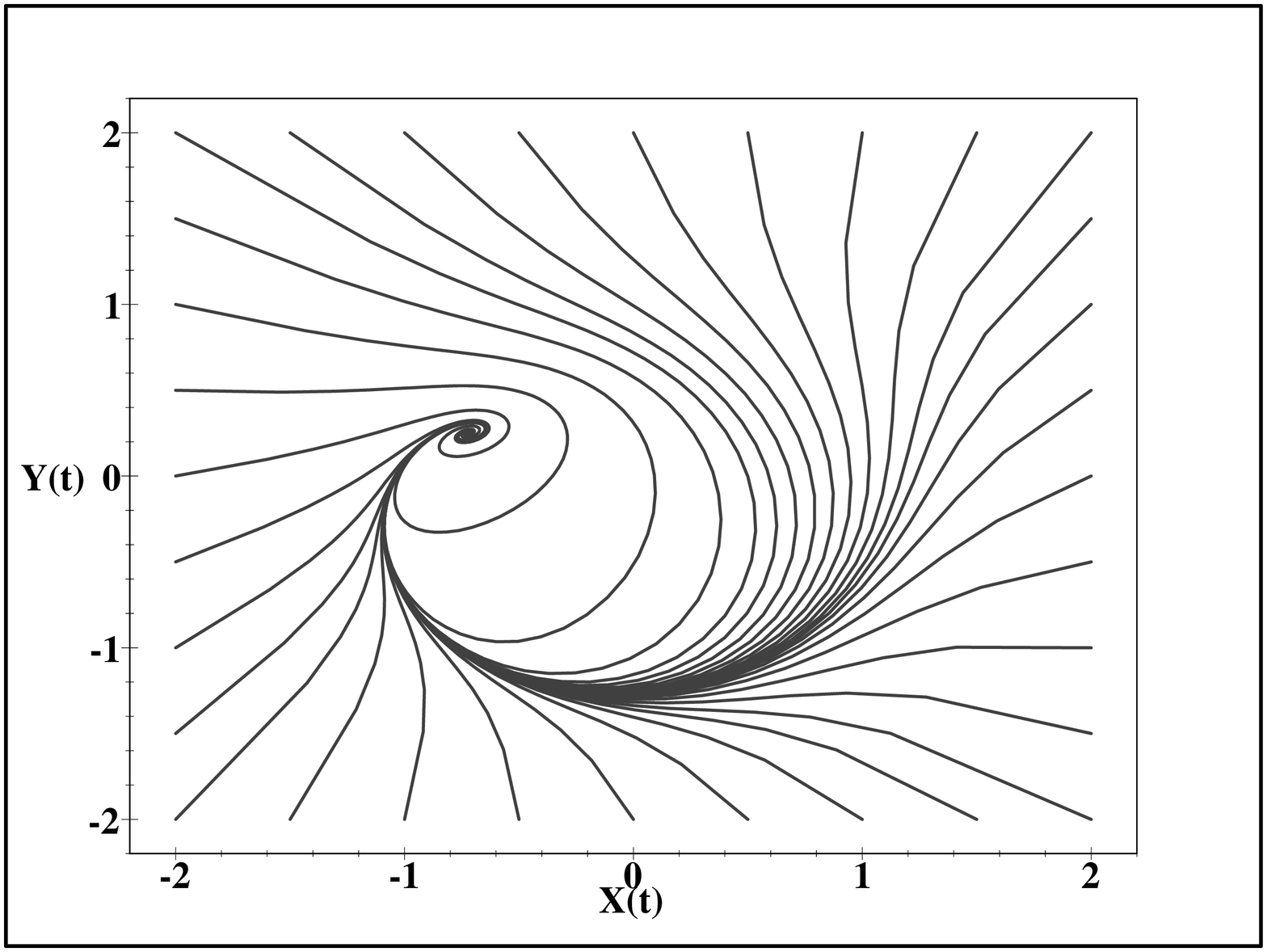}}
\rotatebox{270}{\includegraphics[width=6.7 cm,height=6.7 cm]{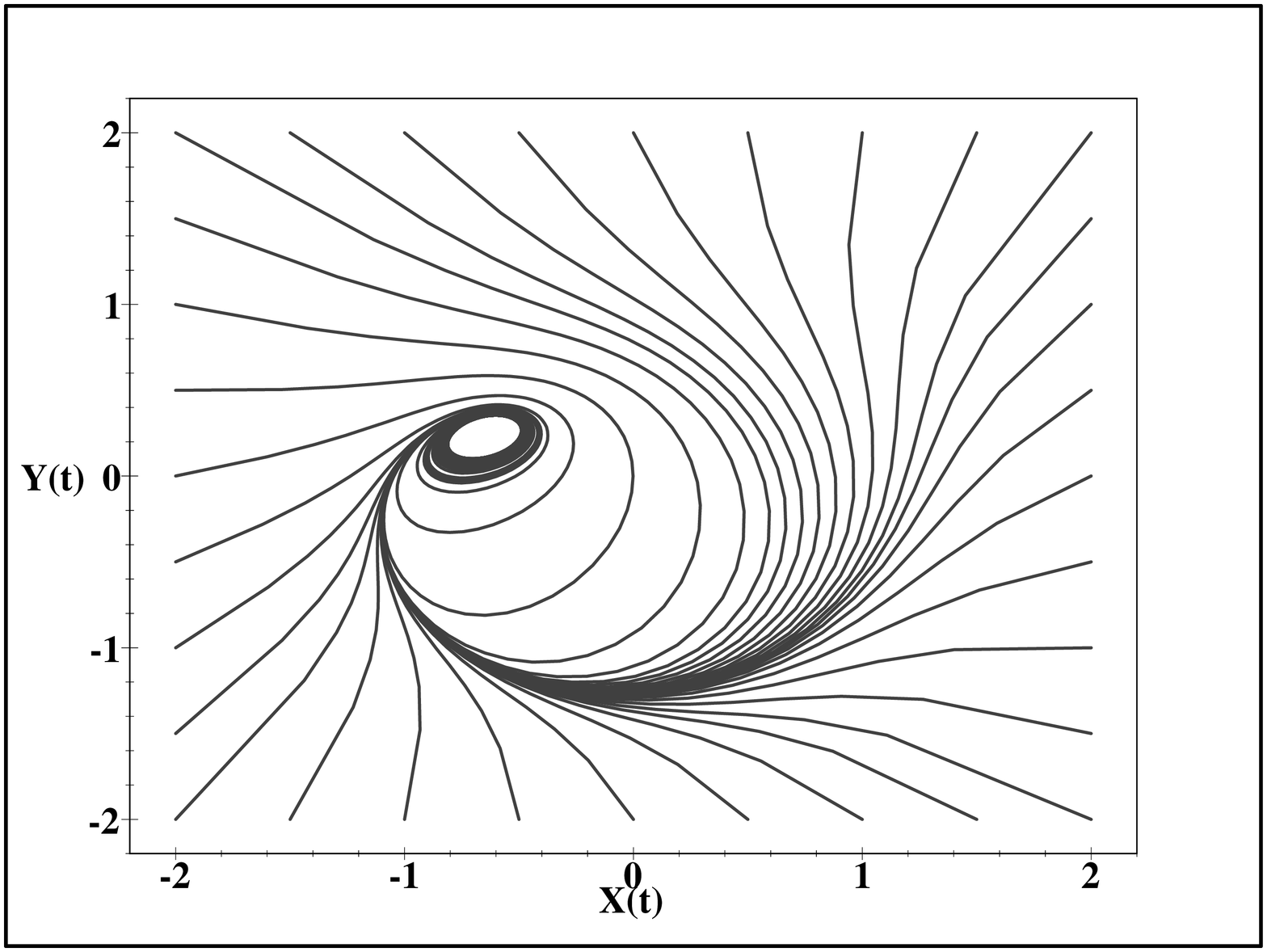}}
\rotatebox{270}{\includegraphics[width=6.7 cm,height=6.7 cm]{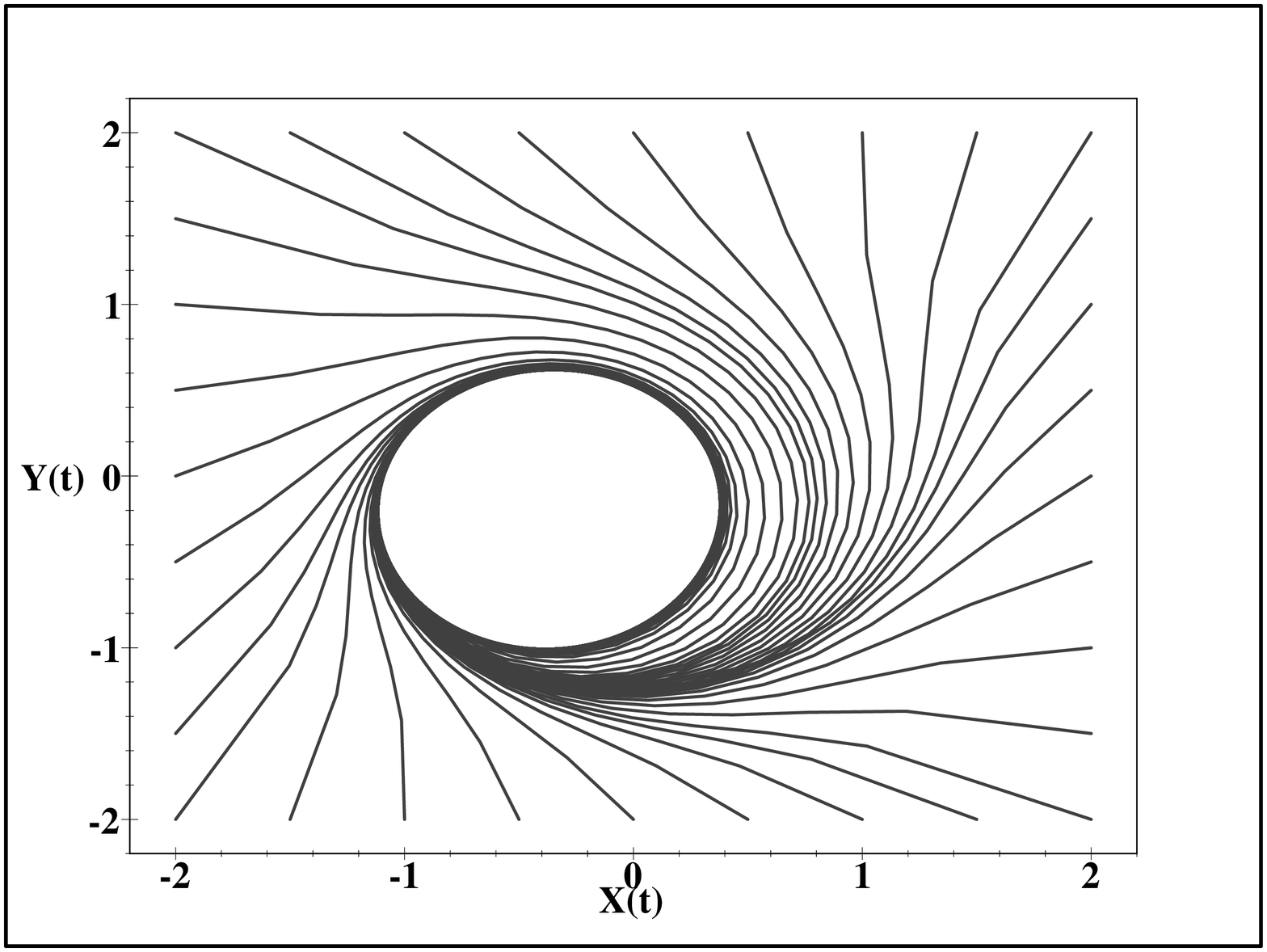}}
\caption{\label{fig4b} 
Approaching to the bifurcation point $\alpha=\alpha_0=1.51...$ 
and transformation to the limit cycle of
solution of forced FGL equation for isochronous case
with fixed wave number $K=0.1$ is represented by real 
$X(K,t)$ and imaginary $Y(K,t)$ parts of $A(K,t)$.
The plots for orders 
$\alpha=2.00$,  $\alpha=1.60$, $\alpha=1.54$, 
$\alpha=1.52$, $\alpha=1.50$, $\alpha=1.40$.
}
\end{figure}

The oscillator medium can be characterized by a single generalized phase 
variable (\ref{A10d}). We can rewrite (\ref{A10d}) as
\be \label{phiXY}
\phi(X,Y)=\arctan (Y/X) -\frac{b}{2} \ln (X^2+Y^2) , 
\ee
where $X$ and $Y$ are real and imaginary parts of $A(K,t)$.
For $E=0$, the phase rotates uniformly
\be
\frac{d}{dt} \phi= - \omega_{\alpha}(K)= a-gc|K|^{\alpha}, 
\ee
where $\omega_{\alpha}(K)$ is gived by (\ref{B7}) with $b=0$, and
can be considered as a frequency of natural oscillations.
For $E \not=0$, Eqs. (\ref{A8e}) and (\ref{phiXY}) give
\be
\frac{d}{dt} \phi= - \omega_{\alpha}(K) -E \cos \phi .
\ee
This equation has an integral of motion.
The integral is
\be 
I=2|\omega^2-E^2|^{-1/2} \arctan \Bigl( 
sgn(\omega-E) \left| \frac{\omega-E}{\omega+E} \right|^{-1/2} 
\tan (\phi(t)/2) \Bigr)+t. 
\ee 
These expressions help to obtain the solution in form (\ref{A9}) for
forced case (\ref{A8e}) keeping the same notations as in (\ref{A9}).
For polar coordinates we get 
\be \label{A10e}
\frac{dR}{dt}= (1-g|K|^{\alpha})R-R^3-E \sin \theta ,
\quad
\frac{d \theta}{dt}= (a-cg|K|^{\alpha})-\frac{E \cos \theta}{R}.
\ee
Numerical solution of (\ref{A10e}) was performed with 
the same parameters as for Eq. (\ref{A8e}), i.e.,
$a=1$, $g=1$, $c=70$, $E=0.9$, $K=0.1$, and $\alpha$ 
within interval $\alpha \in(1,2)$. 
The results are presented in Figs. 2 and 3.

The time evolition of phase $\theta(K,t)$ is given in Fig. 2
for $\alpha=2.00$, $\alpha=1.50$, $\alpha=1.47$, 
$\alpha=1.44$, $\alpha=1.40$, $\alpha=1.30$, $\alpha=1.20$, 
$\alpha=1.10$. The decrease of $\alpha$ from $2$ to $1$ leads
to the oscillations of the phase $\theta (K,t)$ after the Hopf 
bifurcation at $\alpha_0=1.51...$, then 
the amplitude of phase oscillation decreases and
the velocity of phase rotations increases.

The amplitude $R(K,t)$ is shown on Fig. 3 for
$ \alpha= 1.6$,  $\alpha=1.55$, $\alpha=1.55$, $\alpha=1.51$,
$\alpha=1.50$, $\alpha=1.45$, $\alpha=1.2$.  
The appearance of oscillations in the plots
means the loss of synchronization.


\begin{figure}
\centering
\rotatebox{270}{\includegraphics[width=15 cm,height=15 cm]{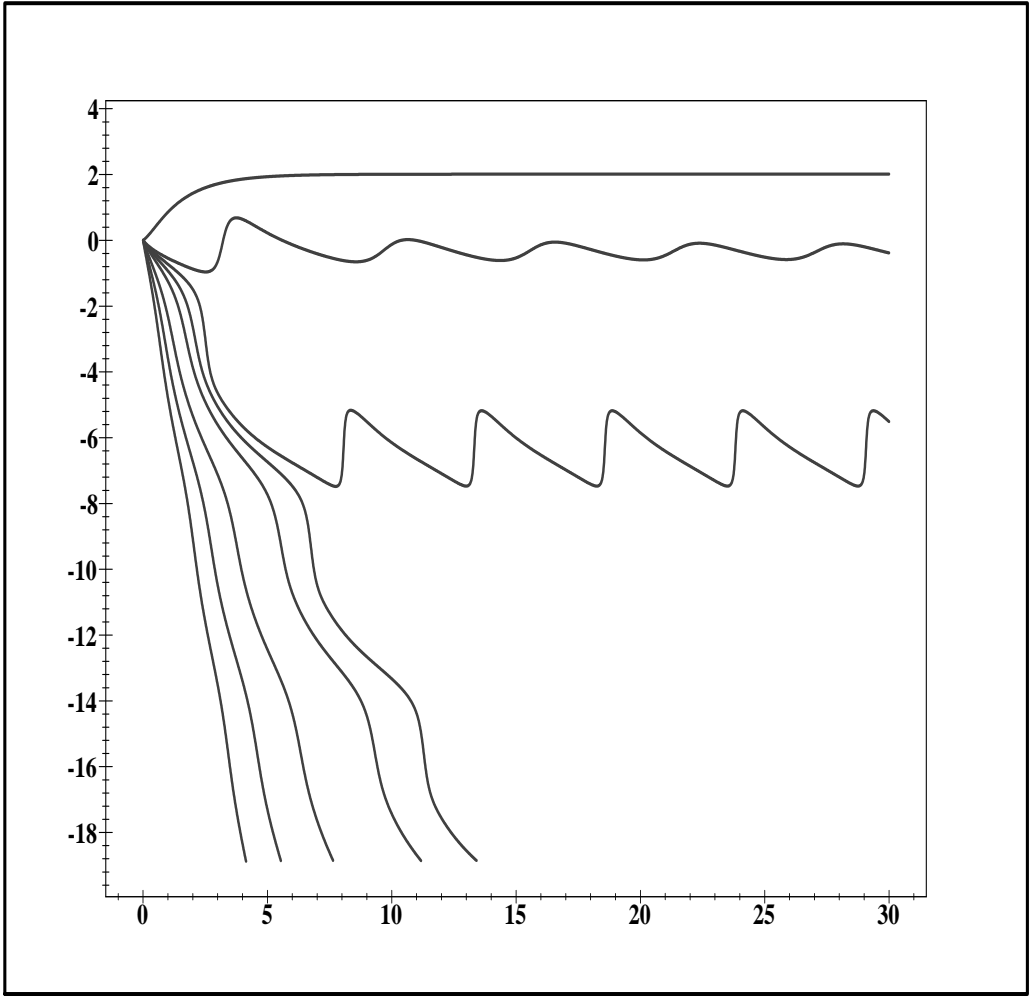}}
\caption{\label{fig2} Phase $\theta(K,t)$ for $K=0.1$ and
$\alpha= 2.00$, $\alpha=1.50$, $\alpha=1.47$, $\alpha=1.44$, 
$\alpha=1.40$, $\alpha=1.30$, $\alpha=1.20$, $\alpha=1.10$.
The decrease of order $\alpha$ corresponds to the clockwise rotation of curves. 
For upper curve $\alpha=2$. For the most vertical curve $\alpha=1.1$. }
\end{figure}

\begin{figure}
\centering
\rotatebox{270}{\includegraphics[width=6.7 cm,height=6.7 cm]{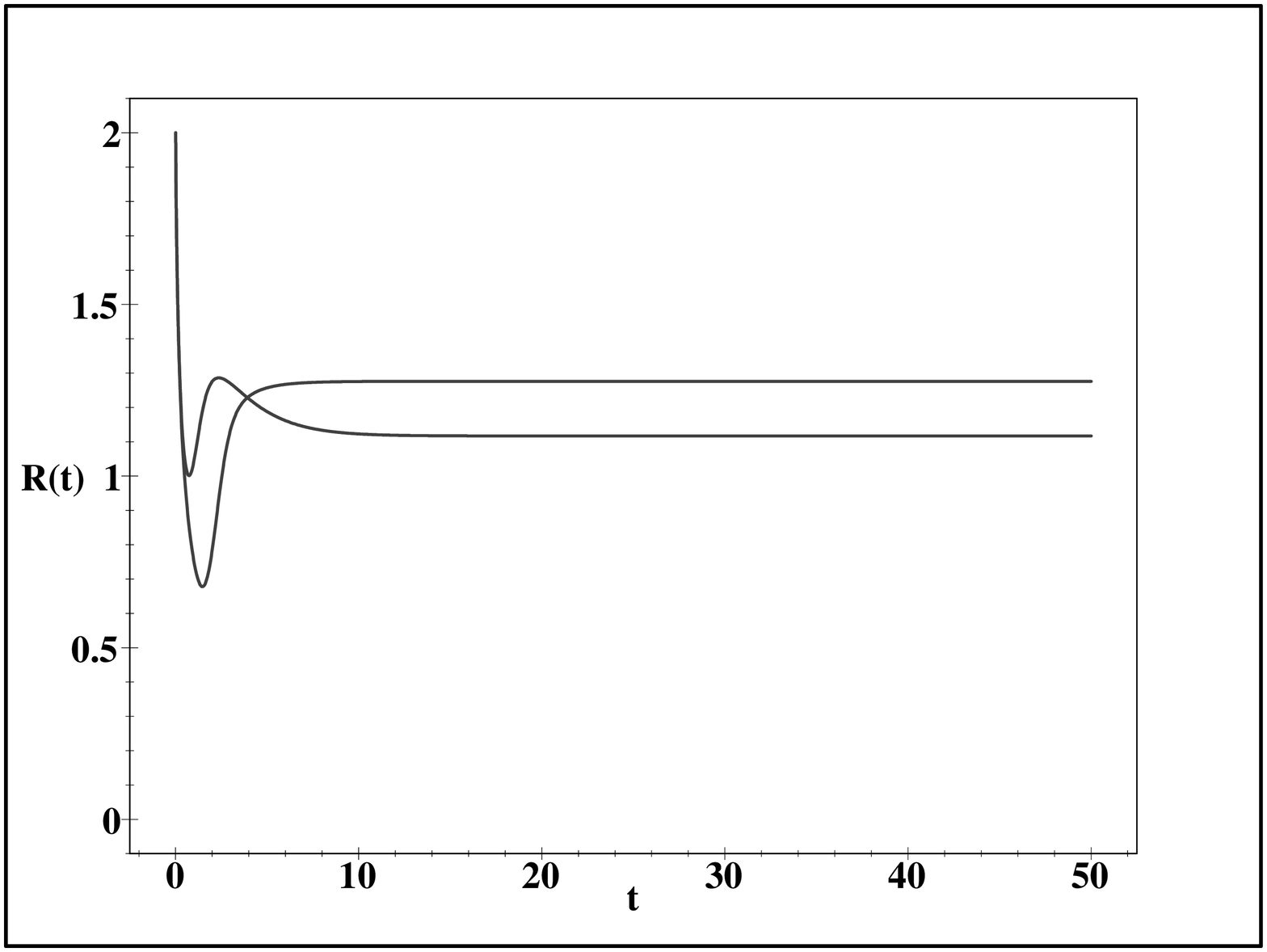}}
\rotatebox{270}{\includegraphics[width=6.7 cm,height=6.7 cm]{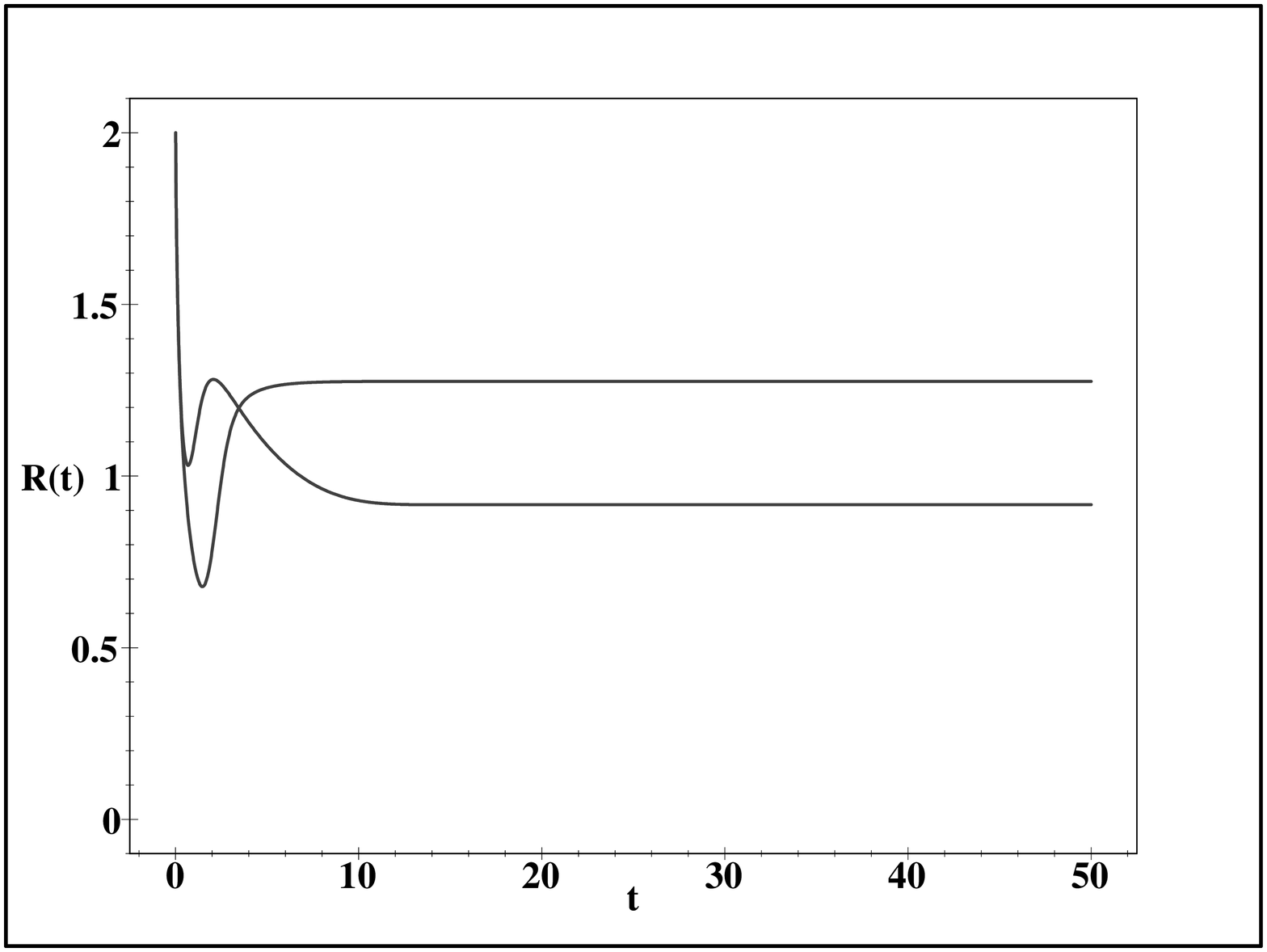}}
\rotatebox{270}{\includegraphics[width=6.7 cm,height=6.7 cm]{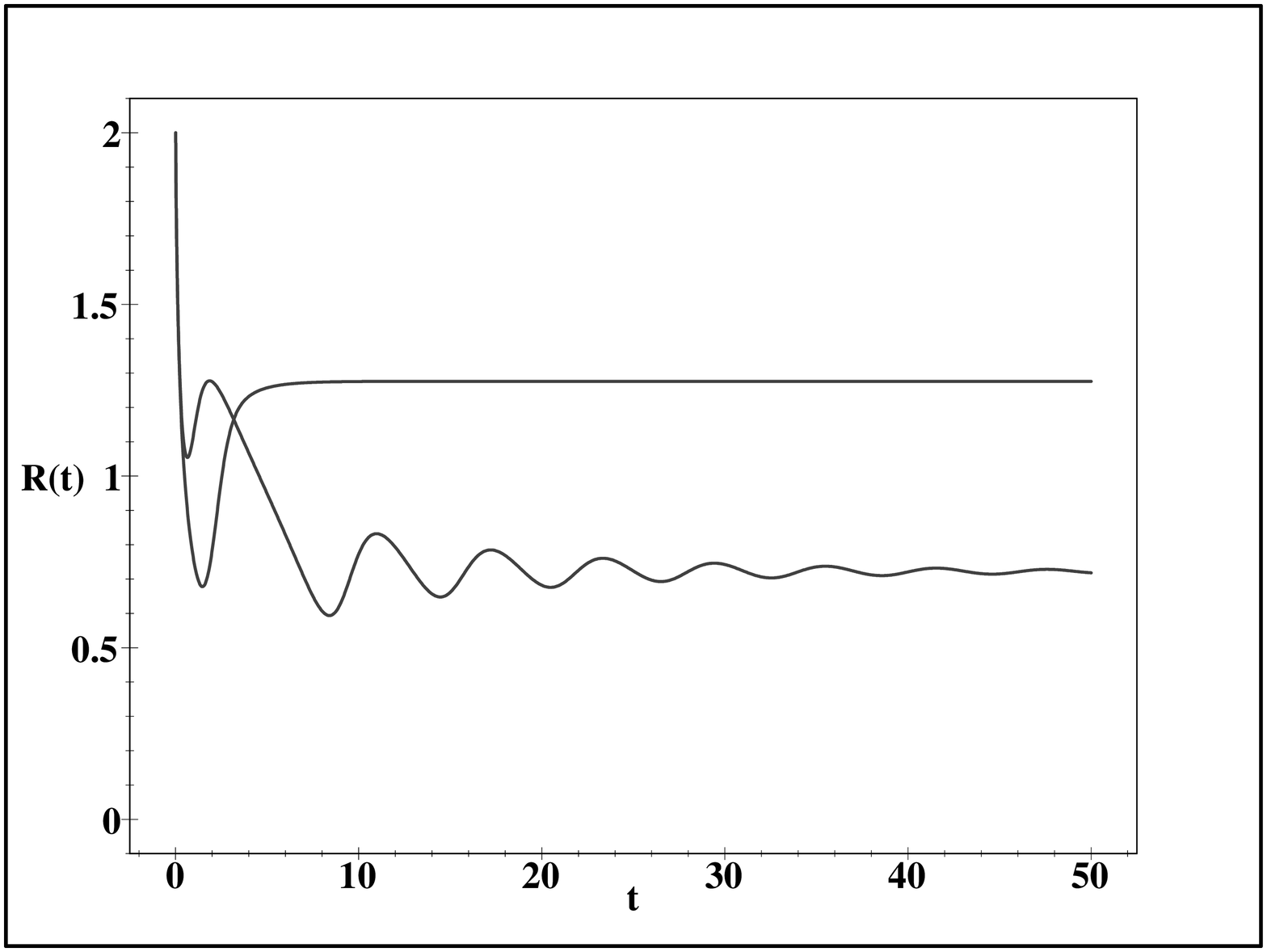}}
\rotatebox{270}{\includegraphics[width=6.7 cm,height=6.7 cm]{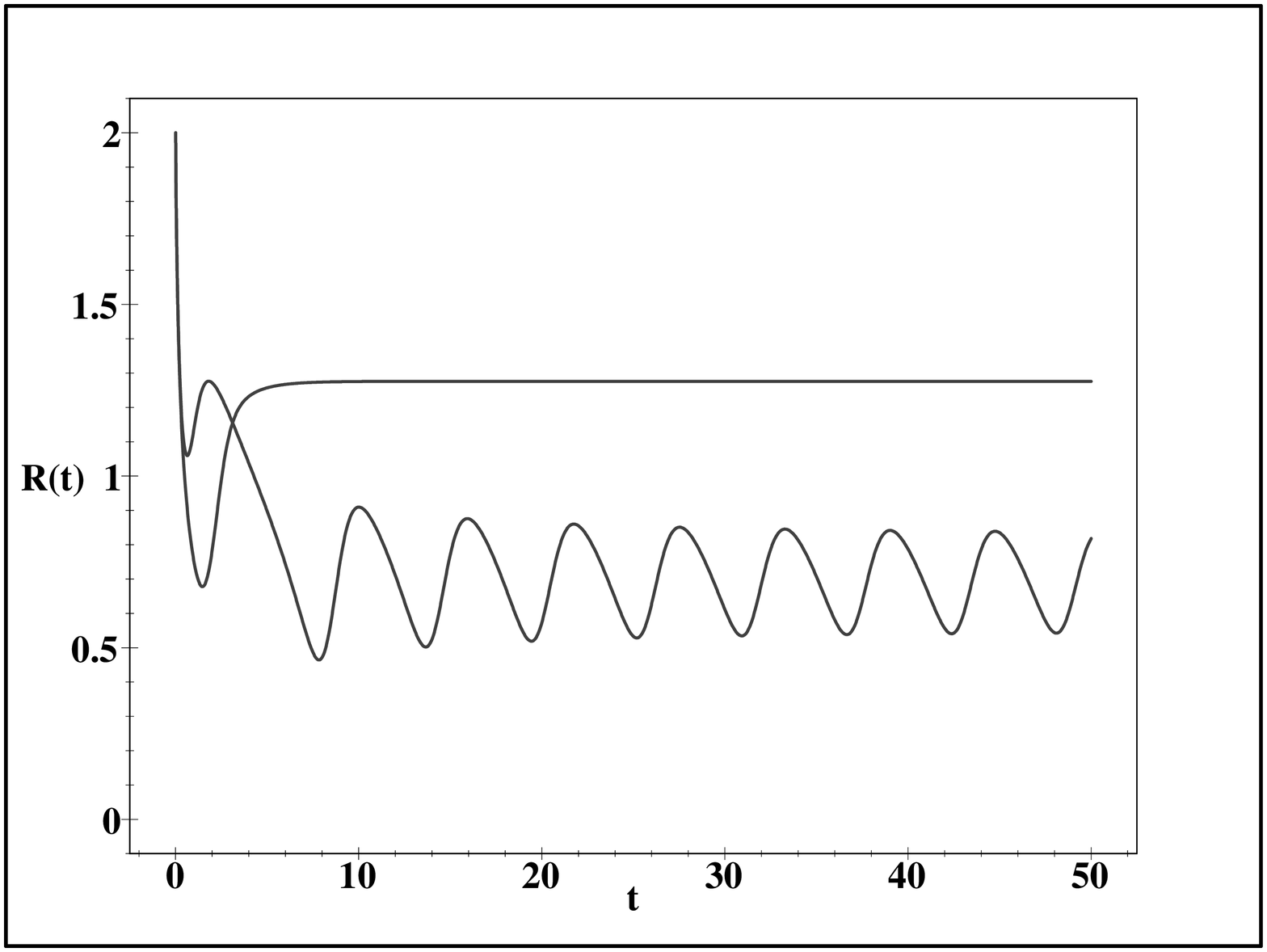}}
\rotatebox{270}{\includegraphics[width=6.7 cm,height=6.7 cm]{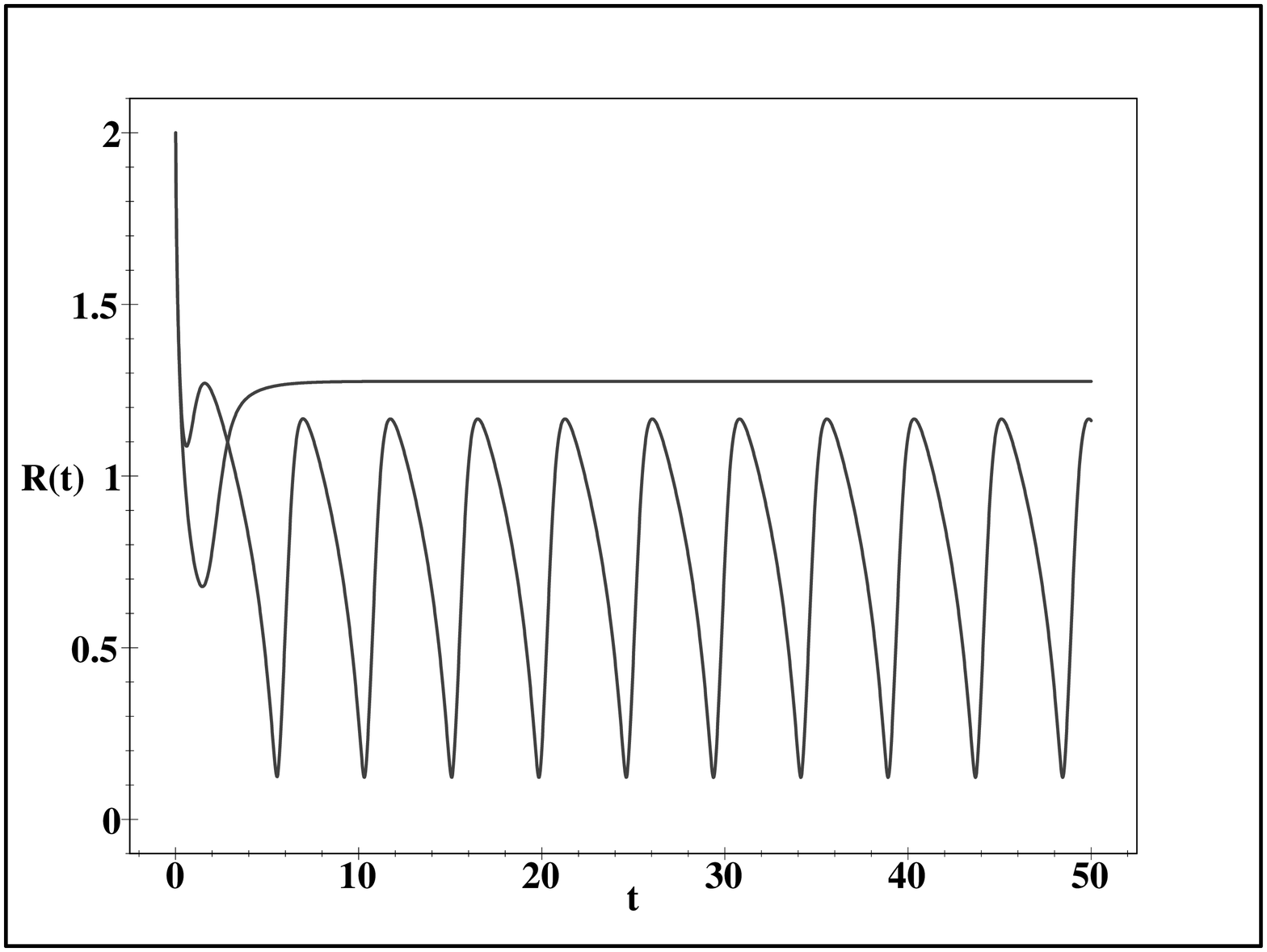}}
\rotatebox{270}{\includegraphics[width=6.7 cm,height=6.7 cm]{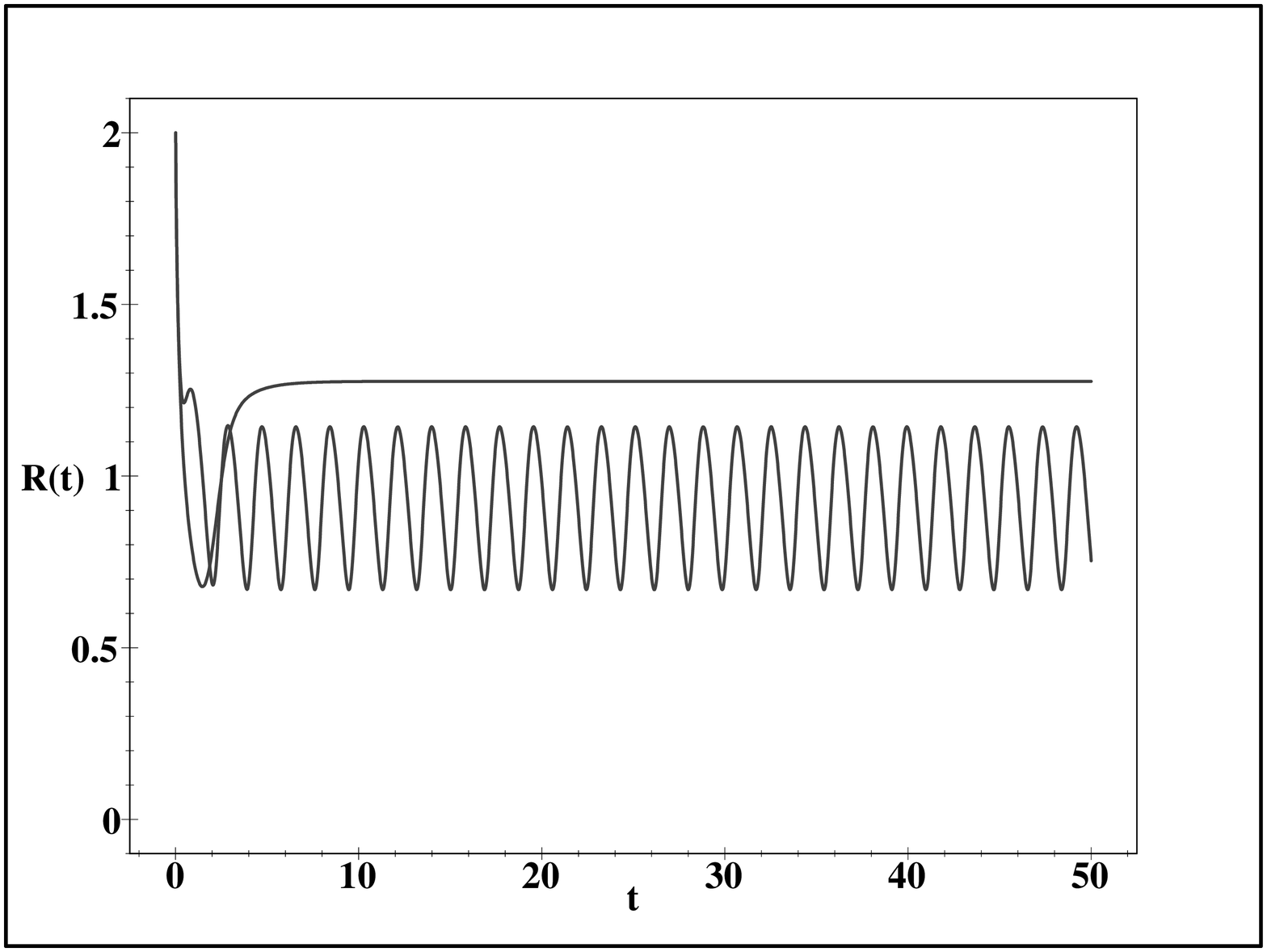}}
\caption{\label{fig3} 
Amplitude $R(K,t)$. The upper curve corresponds to $\alpha=2$ for all plots. 
The lower curves correspond to $\alpha=1.6$, $\alpha=1.55$, $\alpha=1.51$,
$\alpha=1.50$, $\alpha=1.45$, $\alpha=1.2$.  
The appearance of oscillations on the plots
means the loss of synchronization.}
\end{figure}

\section{Space-structures from FGL equation}

In previous sections, we considered mainly time-evolution and 
"time-structures" as solutions for the FGL equation. 
Particularly, synchronization process was an example of the
solution that converged to a time-coherent structure.
Here, we focuse on the space structures for the solution
of FGL equation (\ref{A2}) with $b=c=0$ and the 
constants $a_1$ and $a_2$ ahead of linear term
\be \label{H1}
\frac{\partial}{\partial t}Z=(a_1+ia_2)Z-|Z|^2Z+
g\frac{\partial^{\alpha}}{\partial |x|^{\alpha}} Z .
\ee

Let us seek a particular solution of (\ref{H1}) in the form 
\be \label{H2}
Z(x,t)=R(x,t)e^{i\theta(t)} , \quad R^*(x,t)=R(x,t), 
\quad \theta^*(t)=\theta(t) .
\ee
Substitution of (\ref{H2}) into (\ref{H1}) gives
\be \label{H4}
\frac{\partial}{\partial t}R=
a_1R-R^3- g\frac{\partial^{\alpha}}{\partial |x|^{\alpha}} R , \quad
\frac{\partial}{\partial t}\theta(t)=a_2 .
\ee
Using $\theta(t)=a_2t+\theta(0)$, we arrive to the existence of a 
limit cycle  with $R_0=a^{1/2}_1 $.

A particular solution of (\ref{H4}) in the vicinity 
of the limit cycle can be found as an expansion
\be \label{H9}
R(x,t)=R_0+\varepsilon R_1+\varepsilon^2 R_2+... , \quad
(\varepsilon \ll 1) .
\ee
Zero approximation $R_0=a^{1/2}_1$ satisfies (\ref{H4}) since
${\partial^{\alpha}}/{\partial |x|^{\alpha}} 1=0$,
and for $R_1=R_1(x,t)$, we have
\be \label{HH1}
\frac{\partial}{\partial t}R_1=
-2a_1R_1+ g\frac{\partial^{\alpha}}{\partial |x|^{\alpha}}  R_1 .
\ee

Consider the Cauchy problem for (\ref{HH1}) with initial condition 
$R_1 (x,0)=\varphi(x)$,
and the Green function $G(x,t)$ such that
\be \label{HH3}
R_1(x,t)=\int^{+\infty}_{-\infty} G(x^{\prime},t) 
\varphi(x- x^{\prime}) dx^{\prime} .
\ee
Let us apply Laplace transform for $t$ and Fourier transform for $x$
\be \label{HH4}
\tilde G(k,s)=\int^{\infty}_0 dt 
\int^{+\infty}_{-\infty} dx \ e^{-st+ikx} G(x,t) .
\ee
Applying (\ref{HH4}) to (\ref{HH1}), we obtain
\be  \label{HH8}
\tilde G(k,s)=\frac{1}{s+2a_1+ g |k|^{\alpha} }.
\ee
Let us first invert the Laplace transform in (\ref{HH8}).
Then, the Fourier transform of the Green function: 
\be
\hat G(k,t) = \int^{+\infty}_{-\infty} dx \ e^{ikx} G(x,t)
=e^{ -(2a_1+g|k|^{\alpha}) t }=e^{-2a_1t} e^{ -g|k|^{\alpha} t} .
\ee
As the result, we get
\be \label{HH11}
G(x,t)=(gt)^{-1/\alpha} e^{-2a_1t}  L_{\alpha} ( x (gt)^{-1/\alpha} ) .
\ee 
where 
\be
L_{\alpha}(x)=\frac{1}{2\pi} \int^{+\infty}_{-\infty} dk \  
e^{-ikx} e^{-a|k|^{\alpha}} 
\ee
is the Levy stable p.d.f. \cite{Feller}.

As an example, for $\alpha=1$ we have
the Cauchy distribution with respect to coordinate
\be \label{HHH2}
G(x,t)=\frac{1}{\pi} 
\frac{ (gt)^{-1} e^{-2a_1t}}{x^2 (gt)^{-2} +1}  .
\ee 
For $\alpha=2$, we get the Gauss distribution: 
\be \label{HHH4}
G(x,t)=(gt)^{-1/2} e^{-2a_1t} \frac{1}{2\sqrt{\pi}} 
e^{-x^2 /(4gt) } .
\ee 

For $1< \alpha \le 2$ the function $L_{\alpha}(x)$ can be presented  
as the convergent expansion
\be
L_{\alpha}(x)=-\frac{1}{\pi x} \sum^{\infty}_{n=1} 
(-x)^n \frac{\Gamma(1+n/\alpha)}{n!} \sin (n \pi/2) .
\ee
The asymptotic ($x \rightarrow \infty$, $1<\alpha<2$) is given by
\be
L_{\alpha}(x) \sim -\frac{1}{\pi x} \sum^{\infty}_{n=1} 
(-1)^n x^{-n \alpha} \frac{\Gamma(1+n \alpha)}{n!} \sin (n \pi/2)  , 
\quad x \rightarrow \infty ,
\ee
with the leading term: 
$L_{\alpha}(x) \sim \pi^{-1} \Gamma(1+ \alpha) x^{-\alpha-1}$, 
$(x \rightarrow \infty)$.

As the result, the solution of (\ref{H1}) is
\be \label{SolZxt}
Z(x,t)=e^{i(a_2t+\theta(0))} \left( a^{1/2}_1+ \varepsilon 
(gt)^{-1/\alpha} e^{-2a_1t} \int^{+\infty}_{-\infty} 
L_{\alpha} ( x^{\prime} (gt)^{-1/\alpha} ) \
\varphi(x- x^{\prime}) dx^{\prime} +O(\varepsilon^2) \right) .
\ee
This solution can be considered as a space-time synchronization 
in the oscillatory medium with long-range interaction 
decreasing as $|x|^{-(\alpha+1)}$.

For $\varphi(x)=\delta(x-x_0)$, solution (\ref{SolZxt}) has the form
\be
Z(x,t)=e^{i(a_2t+\theta(0))} \Bigl( 
a^{1/2}_1 + \varepsilon (gt)^{-1/\alpha} e^{-2a_1t}  
L_{\alpha} ( (x-x_0) (gt)^{-1/\alpha} ) +O(\varepsilon^2) \Bigr) ,
\ee
and the asymptotic is
\be
Z(x,t)=e^{i(a_2t+\theta(0))} \Bigl( 
a^{1/2}_1 + \varepsilon gt e^{-2a_1t} \pi^{-1} 
\Gamma(1+\alpha) (x-x_0)^{-\alpha-1} 
+O(\varepsilon^2) \Bigr) , \quad x \rightarrow \infty .
\ee
This solution shows that the long-wave modes approach 
the limit cycle exponentially with time.
For $t=1/(2a_1)$, we have the maximum of $|Z(x,t)|$ with respect to time: 
\be
\max_{t>0} |Z(x,t)|=
a^{1/2}_1 + \varepsilon g  \frac{\Gamma(1+\alpha)}{2\pi e} (x-x_0)^{-\alpha-1} 
+O(\varepsilon^2) .
\ee
As the result, we have the power law decay with respect to the coordinate
for the space structures near the limit cycle $|Z|=a^{1/2}_1 $.

\section{Conclusion}

One-dimensional chain of interacting objects, say oscillators, 
can be considered as a benchmark for numerous applications 
in physics, chemistry, biology, etc.
All considered models were related mainly to the oscillating objects
with long-range power wise interaction, i.e., with
forces proportional to $1/|n-m|^s$ and $2<s<3$. 
A remarkable feature of this interaction is 
a possibility to replace the set of 
coupled individual oscillator equations 
into the continuous medium equation
with fractional space derivative of 
order $\alpha=s-1$, where $0<\alpha<2$, $\alpha\not=1$.
Such transformation is an approximation and 
it appears in the infrared limit for wave number $k \rightarrow 0$.
This limit helps to consider different models and related phenomena
in a unified way applying
different tools of fractional calculus.

A nontrivial example of general property of fractional linear equation 
is its solution with a power wise decay along the space coordinate.
From the physical point of view that means a new type of 
space structures or coherent structures.
The scheme of equations with fractional derivatives
includes either effect of synchronization \cite{Pik1},
breathers \cite{Br3,Br4}, fractional kinetics \cite{Zaslavsky1},
and others.

Discrete breathers are periodic space localized oscillations 
that arise in discrete and continuous nonlinear systems. 
Their existence was proved in Ref. \cite{Br1}. 
Discrete breathers have been widely studied in systems 
with short-range interactions (for a review, see \cite{Br2,Br3}). 
Energy and decay properties of discrete breathers in systems with long-range 
interactions have also been studied in the framework of the Klein-Gordon 
\cite{Br4,Br5}, and the discrete nonlinear Schr\"odinger equations \cite{Br6}.
Therefore, it is interesting to consider breathers solution 
in systems with long-range interactions in infrared approximation.

We also assume that the suggested replacement of the 
equations of interacting oscillators by the continuous medium
equation can be used for improvement of simulations 
for equations with fractional derivatives.

\section*{Acknowledgments}

We are thankful to N. Laskin for  useful discussions and comments.
This work was supported by the Office of Naval Research,
Grant No. N00014-02-1-0056, the U.S. Department
of Energy Grant No. DE-FG02-92ER54184, and the NSF
Grant No. DMS-0417800. 
V.E.T. thanks the Courant Institute of Mathematical Sciences
for support and kind hospitality.


\end{document}